*Article*

# Computational Techniques Enabling the Perception of Virtual Images Exclusive to the Retinal Afterimage

Staas de Jong [1,*]  and Gerrit van der Veer [2]

1 Honours Academy, Universiteit Leiden, 2311 EZ Leiden, The Netherlands
2 Multimedia & Cultuur, Vrije Universiteit, 1081 HV Amsterdam, The Netherlands
* Correspondence: apajong@xs4all.nl

**Abstract:** The retinal afterimage is a widely known effect in the human visual system, which has been studied and used in the context of a number of major art movements. Therefore, when considering the general role of computation in the visual arts, this begs the question whether this effect, too, may be induced using partly automated techniques. If so, it may become a computationally controllable ingredient of (interactive) visual art, and thus take its place among the many other aspects of visual perception which already have preceded it in this sense. The present moment provides additional inspiration to lay the groundwork for extending computer graphics in general with the retinal afterimage: Historically, we are in a phase where some head-mounted stereoscopic AR/VR technologies are now providing eye tracking by default, thereby allowing realtime monitoring of the processes of visual fixation that can induce the retinal afterimage. A logical starting point for general investigation is then shape display via the retinal afterimage, since shape recognition lends itself well to unambiguous reporting. Shape recognition, however, may also occur due to normal vision, which happens simultaneously. Carefully and rigorously excluding this possibility, we develop computational techniques enabling shape display exclusive to the retinal afterimage.

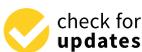

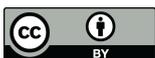

## 1. Introduction

### 1.1. Automated Computation Facilitates Artistic Control over Perceived Visual Complexity

When aiming to broadly compare processes of artistic creation in the visual arts before and after the advent of digital computers, some useful examples to think of may be those where the people involved are graphical artists, and the outcome of the process is the representation of moving 3D imagery. Concretely, we might think of some favorite recent example of computer-generated imagery—be it a cinematically released animated movie, a scientific visual simulation, or a work of visual art completely different altogether—and then try to imagine its accurate reproduction, in all its visual complexity, using only the tools and practices from the era before electronic digital computers became available. Clearly, typically, the nature of the unautomated, manual processes involved in this would now make completion of the work, if at all possible, at the very least much more time-consuming.

This illustrates a broader point, not limited to a specific kind of artwork or aspect of visual perception: A visual artist, by choosing to produce some type of visual effect using partly automated means, may create works based on it that offer a controlled visual complexity which would not be realizable, within the same time frame, if using only unautomated means.

This fundamental advantage has, however, not yet been extended to all aspects of visual perception where it might apply. Here, we will consider the question whether this could be done for the visual phenomenon often known as the *retinal afterimage*.

*1.2. The Retinal Afterimage and Its Use in Techniques for the Visual Arts*

Humans perceive light via the retina, the layer inside of the eye containing light-sensitive cells. The retinal afterimage is a well-known phenomenon in the human visual system, where the preceding exposure to light influences its ongoing perception. The cause for this is thought to be *adaptation*: While, say, staring at a white background with a red figure on it, the retinal cells will adapt to the incoming light, with those cells that respond to green and blue light adapting to relatively lower light levels if they are being exposed to the figure shape. If, after some time, the gaze is then suddenly directed to an entirely white area, the cells adapted to the red shape will respond to light that remains at a constant intensity in the red frequency range, but that has just increased in the green and blue ranges. This can then cause the figure to be perceived again, but now in a color that is complementary to the original red.

To actually perform the above example is to experience the classic afterimage effect, described as a part of human color perception by the opponent-process model [1–3].

Certainly during relatively recent history, various groups of visual artists have developed and used techniques that consciously incorporated the retinal afterimage.

The 19th century, for example, saw the rise of the *Neo-impressionist* movement in painting. Here, Georges Seurat was the seminal figure, for the movement as well as the related techniques of pointillism and divisionism. As a basis for his own work, Seurat used color theories developed and advocated by Michel Eugène Chevreul and other contemporary scientists [4,5]. These color theories were concerned with various observable types of contrast, as well as their presence and use in the visual arts. The retinal afterimage was considered as a part of this, discussed in [6] as "successive" and "mixed contrast" (for a translation, see [7]). Motivated by a desire to achieve maximally brilliant colors, Neo-impressionism took the view that colors should not be mixed as paints on the painter's palette. Instead, color mixing should occur only during the process of viewing [8]. To do this, *divisionism* was developed: a technique where paint is applied in separate dots, in such a way that where adjacent, chosen colors pair up as complementary. The idea is then that for areas where colors are paired in this way, the viewer may perceive a given color in a more bright or intense way, due to retinal afterimages induced by the respective complementary color. This assumption is still held in recent educational materials [9].

Moving to the 20th century, during its second half, a different art movement also gave special attention to the process of viewing, while exploring and employing many illusory phenomena in the human visual system: the Optical art or *Op art* movement [10,11]. The illusory phenomena included the retinal afterimage, for example in works by the painters Bridget Riley [12], Richard Anuszkiewicz [13], and Larry Poons [14]. More generally, the retinal afterimage is part of the techniques associated with Op art [15]. Where countless demonstrations, from Goethe [16] to the 21st century [17], have used afterimage shape figuratively, Op art has often used it to show abstract forms.

Also, going back to earlier in the 20th century, one of the people teaching at Germany's original *Bauhaus* school was Josef Albers. Over time, his artistic and educational work came to have a strong influence in design and visual art (see e.g., [18]). In Albers' educational materials, the retinal afterimage is presented as something to generally take into consideration when using color in design and visual art, as a fundamental phenomenon in human color perception [19].

Finally, an influential figure in film was Stan Brakhage, who produced his experimental oeuvre partly by painting directly on the frames of analog film strips. At one point, he considered afterimage colors the only true colors [20].

*1.3. Two-Dimensional Shape Display as a Test Case for Computing the Retinal Afterimage*

Could we partly automate techniques for inducing the retinal afterimage? The above discussion has only further motivated a return to this question, having made concrete how, over time, various individuals and movements in visual art have considered the retinal afterimage to be a significant phenomenon—one to be studied as well as incorporated into the artistic techniques being developed and used.

To then make the central question itself and the process of finding an answer to it more concrete, we should first make a basic choice regarding the algorithms to be developed: What specific aspects of the afterimage should these offer to the visual artist for specification? This could include for example shape, color, and textural effects. Out of these, we choose to pursue 2D shape display first, because when testing new techniques with test subjects, the recognition of shapes seems especially well-suited to unambiguous reporting.

*1.4. Fundamental Problem: Avoiding Shape Recognition Outside the Afterimage*

Our choice for 2D shape display via the retinal afterimage does present a fundamental problem however, which is directly illustrated by the afterimages that can be induced using Figures 1a and 1b. Already beforehand, the shapes of sinusoid bands can be recognized in the first Figure, and the representation of a face in the second. Because of the repeating symmetry of Figure 1a, its afterimage will show the already-seen sinusoid shapes, in what seems to be a nearly identical, but shifted version of the original image. The afterimage induced by Figure 1b, on the other hand, will resemble its predecessor in a much less obvious way, briefly showing a representation of the face now inverted back to a positive. This diminished resemblance is reflected in the greater element of surprise associated with perceiving the second afterimage.

As these examples clearly show, the contents of the retinal afterimage may, to a varying extent, already be recognizable via normal viewing of the imagery used to induce the effect. Because of this, any technique claiming to offer 2D shape display via the retinal afterimage must carefully and rigorously exclude the possibility that it is normal vision that is actually (in part) responsible for shape recognition by the viewer.

In order to keep clear of this kind of false positive, we will require that outside of the afterimage effect, results for shape recognition will be explicitly negative.

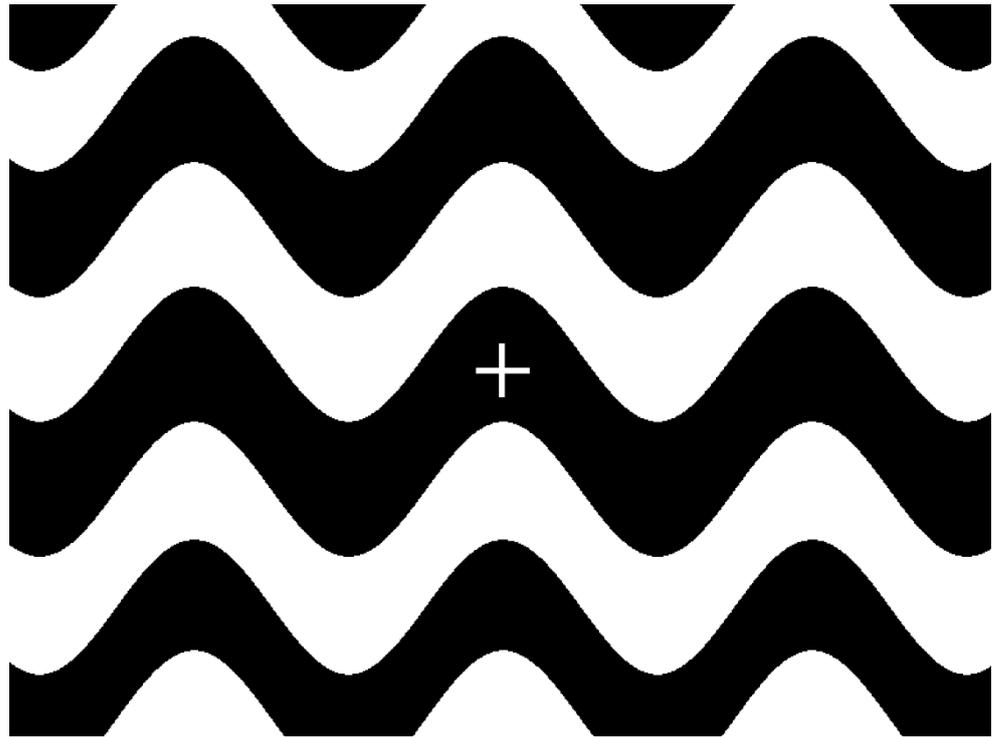

**Figure 1a.** Example afterimage shapes. For viewing, first enlarge this image, and increase its visual contrast as much as is possible and comfortable. Then, from close by, focus on the crosshair in the middle for about one minute. After this, close your eyes. Briefly, an afterimage will appear. Subsequent blinking may bring it back, as it becomes less and less distinct.

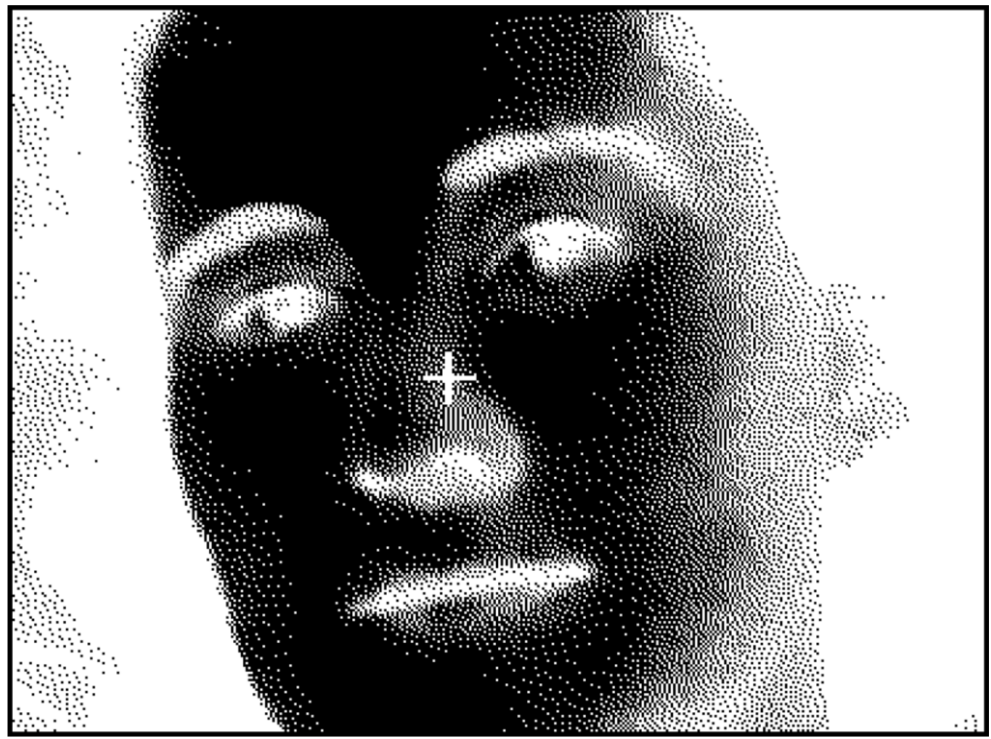

**Figure 1b.** A second example afterimage shape. (See Figure 1a for viewing directions.)

## 2. Materials & Methods: Computing Shapes Exclusive to the Retinal Afterimage

Besides requiring explicitly negative results for shape recognition outside of the afterimage effect, we will limit our automated approach to the greyscale case, and not use different perceived light intensities for different types of retinal color receptors. Also, we will induce the retinal afterimage using the minimum amount of images: one *bias image* for retinal adaptation, followed by one *trigger image* triggering the effect. Having two such greyscale images produce an afterimage showing the specified 2D shape, while this shape is not recognizable in either of the images separately, then corresponds to visualization exclusive to the retinal afterimage.

Below, the construction of a method for computing 2D shapes in the retinal afterimage will be discussed. This will be illustrated by figures in the text, and by video examples that can be found in the Electronic Appendix. All development was done using the TFT LCD display of a Packard Bell R3450 laptop computer, which has a resolution of 98 dpi. This display was used during daytime (set to maximum brightness), in an indoors setting otherwise without artificial lighting. Image sequences were viewed from a distance of approximately 60 cm, chosen as a typical and comfortable viewing distance. Bias images were displayed for 20 s, and immediately followed by their trigger images.

Software implementing the method has also been included in the Electronic Appendix, allowing direct experimentation with various aspects of the method. These include input patterns, visualization parameters, and rule sets, discussed below. It may be necessary to use the software to recreate video examples, for playback on other display devices than the one used here: Differences in color gamut between various display types and technologies can be considerable, and may initially prohibit an effective reproduction of the greyscale intensities used. The solution in such cases would be to repeat construction as described, but using the specific display in question.

### 2.1. Controlling Induced Afterimage Intensity: Visual Fixation

Our basic means for facilitating well-defined, predictable retinal adaptation will be to have the viewer fixate visually. This will be done much like in the traditional examples that employ shapes put on paper. However, unlike those examples, subsequent eye movement, necessary to view the retinal afterimage, will be eliminated from the process. Instead, once retinal adaptation has occurred, the digital display technology used will seamlessly switch the image already being looked at to another one, and thereby trigger the afterimage effect. Then, within the images used, a specific crosshair shape will be placed as a visual aid for fixating. It will have black and relatively thick edges, to make it easier for the viewer to focus on its center regardless of the varying contents of the images around it. The center itself will be defined by the intersection of two hairlines, drawn in green to further stand out from the greyscale imagery. The width of the hairlines will be one pixel, and their length 40 pixels. The resulting crosshair shape is shown in Figure 2.

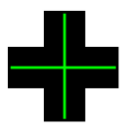

**Figure 2.** The crosshair for visual fixation.

During a typical image sequence, the viewer will be asked to focus constantly on the crosshair's center, which in practice will probably mean constantly correcting for small fixational errors. This can be facilitated by supporting the head with both hands while the elbows are resting on some surface, as suggested in [21].

### 2.2. Simultaneously Inducing Different Afterimage Intensities: A Rasterization Method

To display 2D shapes, the retinal afterimage will have to create the simultaneous perception of different light intensities at different locations. To do this, we will simply divide the bias and trigger images into square areas of *m* × *m* pixels. Through visual fixation, each square in a bias image will correspond to a square in the following trigger image, creating a separate sequence of greyscale intensities. We will use a total image size of 800 × 600 pixels, which allows full-size playback on many different types of displays.

If we try this out for *m* = 50 however, using a chequered black and white pattern (see Figure 3, to the left) followed by a similar image where black is replaced by a light grey (of 90% intensity), we find that unintended lighter and darker shades appear along the square edges in the trigger image. This can be seen in the first sequence of Video Example I.

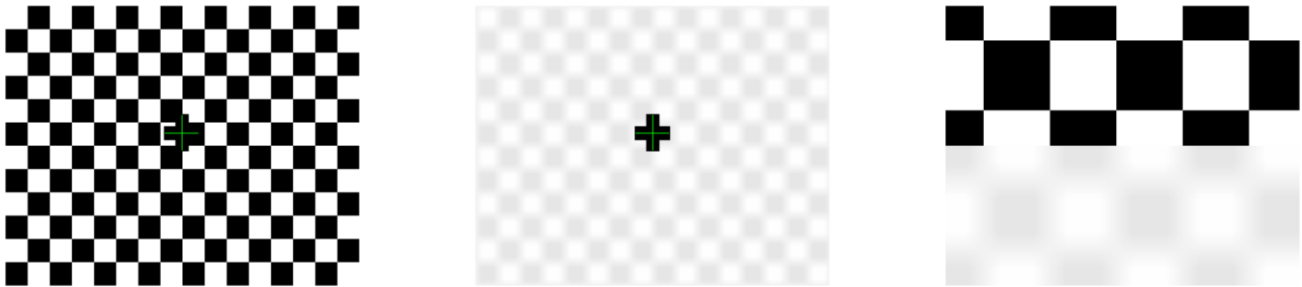

**Figure 3.** Example sequence. Left: cropped middle section, around the crosshair, of a bias image. Middle: the corresponding section of a trigger image. Right: a more detailed comparison, illustrating convolution of the trigger image (*n* = 13).

A possible explanation is that we do not succeed in projecting adaptation patterns precisely over those due to the trigger image. This is not surprising when recalling the afterimages produced before, which were not exact inversions, but blurred variations of the original images. In addition, the already blurred pattern of retinal adaptation cannot be placed exactly and steadily over the trigger image, because of continuous, involuntary small eye movements [22].

We may, however, try to alleviate the effects of this, by blurring the edges in the trigger image. The idea here is that the middle sections of squares, which do overlap, produce the desired afterimage intensities; while the borders in between those intensities, although unstable, can be made into smooth transitions, by having the light differences resulting from small eye movement be more gradual.

The contents of a pixel matrix, such as a trigger image, can be blurred by convolving them with the contents of another matrix, which specifies how each pixel's greyscale value is to be recomputed as a weighted sum of the values of itself and its neighbours. We will use $n \times n$ convolution matrices, with all elements equal to $1/n^2$, so that in general the replacement of the value $p_{x,y}$ of a pixel at location $(x, y)$ will be the mean

$$\frac{1}{n^2} \sum_{i=0}^{n-1} \sum_{j=0}^{n-1} p_{x-\lfloor n/2 \rfloor+i,\ y-\lfloor n/2 \rfloor+j}$$

(Here, pixels lying outside the pixel matrix are assigned the value of the nearest pixel inside the matrix.) For small *n*, this results in bands of *n* - 1 pixels wide between the square areas, linearly traversing the difference in greyscale intensity. For odd *n*, the middle of such a band will be at the location of the original sharp edge (see Figure 3 for an example). Now, as *n* increases from *n* = 1, the perception of unintended shades along the square edges in the trigger image decreases; around *n* = 13 it may seem largely gone, and replaced by an impression of squares of even intensity, with fuzzy borders between them. This can be seen in the second sequence of Video Example I, where the trigger image has been convolved for *n* = 13.

This sequence also shows another effect, however: squares in the afterimage may sometimes seem to join their neighbours, resulting in areas of even intensity which break the regular chequering pattern. We might have anticipated that different sequences of greyscale intensities could be used to create similar afterimage intensities, but these irregularities seem surprising in that they show that apparently similar causes—the regular structures in the bias and trigger images—do not always lead to similar results in the afterimage. In [23], a range of potential factors influencing the complete or partial disappearance and reappearance of structured afterimages is examined. In [24], the influence of selective attention in this is highlighted and studied, including an effect of filling-in of enclosed regions. The irregularities that can be observed in the chequering pattern might provide an example of related effects.

Still, it seems possible that these squares with fuzzy borders could be used to construct the display of 2D shapes in the retinal afterimage. For this, it would be desirable to increase shape resolution by decreasing square size $m$. A minimum for this would be $m = n$, since below this value, convolution would not leave the original greyscale intensities of trigger image squares present. Going down from $m = 50$, there initially ($m = 38$, $m = 32$, $m = 25$) is the impression of fuzzy squares as described before—imperfect, but apparently similarly so. For lower values ($m = 22$, $m = 19$, $m = 13$) the squares give an increasingly unstable impression, distorted by large diagonal patterns. The third sequence of Video Example I again shows the chequered sequence, now with square size decreased to $m = 25$ (and $n = 13$ as before).

Above, we have determined $n$ and $m$ only tentatively; the software included in the Electronic Appendix allows free experimentation with both parameters.

*2.3. A Naive but Formal Model of Induced Afterimage Intensity*

In the previous Section, the retinal afterimage of Figure 3 and Video Example I seemed to give an overall impression according with our initial discussion of the afterimage: The light grey squares preceded by black ones appeared to light up, which can be explained by retinal adaptation to lower light levels. In order to realize a general method for 2D shape display, we would like to explore these effects of adaptation in a formal way. We will do this by making a number of naive assumptions, giving us a simple model to work with—explicitly not intended as a model for the retinal afterimage in any general sense. Factors influencing afterimage color perception which will not explicitly be taken into account here yet for example include post-adaptation contour alignment [25] and the presence of induced contrast, both during and after adaptation [26].

Before formulating assumptions, we have to make explicit a distinction between *bias* and *trigger intensity* on the one hand, and *afterimage intensity* on the other. The former will mean actual light levels produced by screen pixels, corresponding to greyscale values stored in display memory. The latter, a perceived intensity, will indicate a scale between dark and light based on the subjective impression created in the viewer.

The first naive assumption we will use, then, is that using one particular sequence of bias and trigger intensities will normally result in the perception of one particular afterimage intensity. Denoting the set of possible pixel display intensities as a finite subset $I \subset [0, 1]$ (where 0 means black and 1 means white), we model afterimage intensity as a real-valued function $f_a: I \times I \to R$ of pixel display intensities (while assuming environment light, exposure times and retinal sensitivity to be constant).

Then, if bias intensity $b$ and trigger intensity $t$ both have the same value for a sequence (as e.g., for the white squares in Figure 3), we denote the resulting afterimage intensity by this value, and assume that changing the bias intensity would cause it to change as well:

$$b = t \Leftrightarrow f_a(b, t) = t$$

From the earlier discussion of retinal adaptation, we would expect such a change in bias intensity to be associated with either a decrease or an increase in afterimage intensity (as e.g., for the light grey squares in Figure 3):

$$b > t \Leftrightarrow f_a(b, t) < t$$
$$b < t \Leftrightarrow f_a(b, t) > t$$

Our final assumption (which will be exemplified by Figure 4) is that one particular trigger intensity will give a darker impression in the afterimage if and only if there has been retinal adaptation to a lighter bias intensity:

$$b_1 > b_2 \Leftrightarrow f_a(b_1, t) < f_a(b_2, t) \tag{1}$$

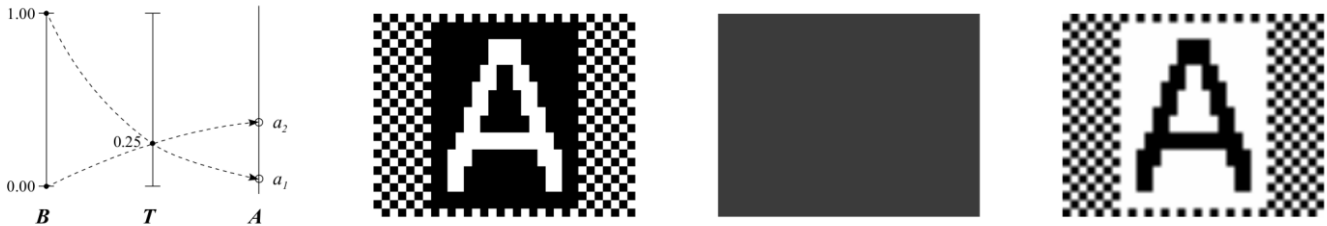

**Figure 4.** Trigger ambiguity. Left: diagram of rule set $f_1$, with arrows linking the bias and trigger intensities that are used to their target afterimage intensities. Middle, left: the bias image from an image sequence generated by rule set $f_1$. Middle, right: the evenly grey trigger image. Right: the target afterimage pattern, with $a_1$ shown as black, and $a_2$ shown as white.

*2.4. Visualization Exclusive to the Retinal Afterimage: Rule Sets*

Having made these assumptions, suppose now we would want to produce afterimages using some set $A = \{a_1, a_2, \ldots, a_n\} \subset R$ of $n > 1$ afterimage intensities, with $a_1 < a_2 < \ldots < a_n$. There would probably be various ways to arrive at such a set, but in any case, we would have to select pairs of bias and trigger intensities with which to produce it. We define an afterimage *rule set* for producing $A$ as a partial function $f_r : I \times I \to A$ which is surjective, so that it produces all of $A$:

$$\forall a \in A \, \exists b, t \in I : f_r(b, t) = a$$

and which respects the afterimage function $f_a$:

$$\forall a \in A : f_r(b, t) = a \Rightarrow f_a(b, t) = a \tag{2}$$

Now, given a rule set and a particular pattern of afterimage intensities to be produced, we can determine the contents for corresponding bias and trigger images by applying the rule set to the pattern. In the following subsections, we will define various types of rule sets (accompanied by concrete examples), with properties in favor of shape display exclusive to the afterimage. For this, it is important to mention first that rule sets will be used non-deterministically: If more than one $(b, t)$ pair may produce a given $a = f_r(b, t)$, one of them will be chosen at random.

2.4.1. Ambiguous Rule Sets

First, we will focus on the subgoal that a pattern of afterimage intensities should not be recognizable in the trigger image used to produce it. A certain way to achieve this would be to use a rule set in which any trigger intensity can lead to any afterimage intensity: We could then freely choose our pattern of trigger intensities, regardless of the afterimage that is to be produced. We say a rule set $f_r$ is *trigger-ambiguous* if:

$$f_r(b, t) = a \Rightarrow \forall a' \in A \, \exists b' \in I : f_r(b', t) = a'$$

The simplest examples of such rule sets would use a single trigger intensity to produce the minimum two afterimage intensities. Consider for example the rule set $f_1$, shown in the diagram of Figure 4. (In the diagram, $B$ and $T$ indicate the subsets of $I$ containing $f_1$'s bias and trigger intensities, respectively. These subsets are formally defined below.) Rule set $f_1$ is based on a dark grey trigger intensity, which is either darkened by a preceding white intensity, giving $a_1 = f_1(1, 0.25)$, or lit up by a preceding black intensity, giving $a_2 = f_1$ (0, 0.25). This will not allow a choice in trigger intensity, but the ambiguity requirement still guarantees that shapes in the afterimage will not be recognizable in the always evenly grey trigger images.

This is illustrated by the image sequence depicted in Figure 4, and demonstrated in Video Example II, which is the result of applying $f_1$ to an example afterimage pattern consisting of a symbol and a simple regular pattern. This sequence also shows how, by generating negative bias images and neutral trigger images, rule set $f_1$ implements the classic afterimage effect.

Since we would like shapes in the afterimage to go unrecognized in the bias image also, we introduce a second concept, analogous to trigger ambiguity. We say a rule set $f_r$ is *bias-ambiguous* if:

$$f_r(b, t) = a \Rightarrow \forall a' \in A \; \exists \, t' \in I : f_r(b, t') = a'$$

Our example for this type of rule set will introduce the simultaneous use of different bias-trigger intensity pairs to produce the same afterimage intensity. The starting point here will be the fact that black and white bias intensities can lead to two distinct afterimage intensities by simply remaining constant. This corresponds to a rule set $f_2$, shown in the diagram of Figure 5, of which the first half is given by $a_1 = f_2(0, 0)$ and $a_2 = f_2(1, 1)$. To satisfy the ambiguity requirement, the second half will then have to allow each bias intensity to lead to both of the afterimage intensities.

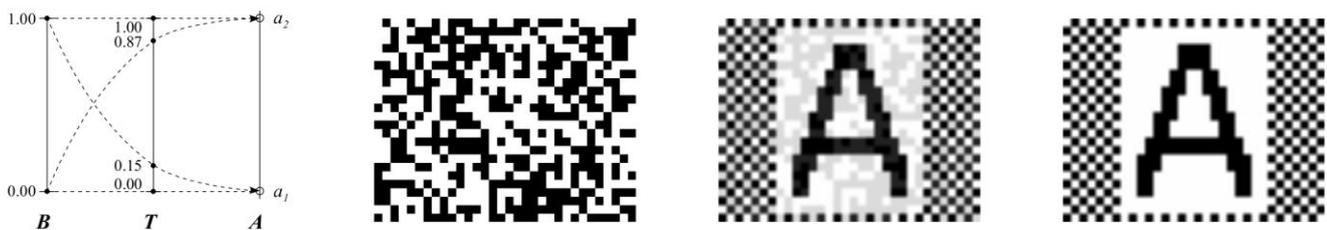

**Figure 5.** Bias ambiguity. Left: diagram of rule set $f_2$, with arrows linking the intensities that are used. Middle, left: the random bias image from an image sequence generated by rule set $f_2$. Middle, right: the slightly distorted trigger image. Right: the target afterimage pattern, with $a_1$ and $a_2$ shown as before.

In the case of black, for example, we have to find a $t$ for which $a_2 = f_2(0, t)$. This can be done by varying $t$ and judging how well the combination of $f_2(0, t)$ and $f_2(1, 1)$, applied to an all-$a_2$ afterimage pattern, succeeds in creating an even impression in the afterimage. Three ranges will then be distinguishable for $t$: a high range where $f_2(0, t)$ will seem lighter than $f_2(1, 1)$; a low range where the opposite is true; and a range inbetween, where the afterimage intensities will seem similar, but still may flicker or otherwise remain visually separate. The most stable results in this middle range seemed to be around $t = 0.87$, so that we will use $a_2 \approx f_2(0, 0.87)$.

Analogous steps taken for the white case, while testing with an all-$a_1$ pattern, then yielded $a_1 \approx f_2(1, 0.15)$, finishing rule set $f_2$.

The next step, of using $f_2$ on a concrete target pattern, will mean that for each afterimage intensity within that pattern, a random choice will be made out of those bias-trigger intensity pairs that could produce it. Overall, the resulting bias image will therefore also be a random black-and-white pattern. This pattern will then by the same process also be present in the trigger image, distorting it in a limited way. This is demonstrated in Figure 5 and Video Example III, which use the same example afterimage pattern as before.

The two examples above have concretely demonstrated how viewers will not be able to recognize the target afterimage pattern where an ambiguity requirement applies—be this in the trigger or bias image, respectively. However, it is also true for each example that in the image where ambiguity did *not* apply, the afterimage shape could easily be recognized. This leads us to wish for a rule set $f_r$ that has both bias *and* trigger ambiguity simultaneously. However, regrettably, such a *fully ambiguous* rule set cannot exist, and we can prove this within our framework.

**Proof.** Suppose a rule set $f_r$ is bias-ambiguous.
　Define the sets $B$ and $T$ of $f_r$'s bias and trigger intensities:
　　$B = \{ b \mid b \in I \land (\exists\, t \in I, a \in A : f_r(b, t) = a) \}$
　　$T = \{ t \mid t \in I \land (\exists\, b \in I, a \in A : f_r(b, t) = a) \}$
　Then choose
　　$b_{max} \in B$ so that $\forall b \in B : b \leq b_{max}$
　　$a_{max} \in A$ so that $\forall a \in A : a \leq a_{max}$
　　$a_{min} \in A$ so that $\forall a \in A : a \geq a_{min}$.
　Due to bias ambiguity, there must exist a $t' \in T$ with
　　$f_r(b_{max}, t') = a_{max}$.
　However, for this $t'$, trigger ambiguity will not hold, because
　　$\neg\, \exists\, b' \in B : f_r(b', t') = a_{min}$.
　　*Proof.* Suppose $\exists\, b' \in B : f_r(b', t') = a_{min}$.
　　　Then $f_a(b', t') < f_a(b_{max}, t')$ by Prop. (2),
　　　since $n > 1$ guarantees $a_{min} < a_{max}$.
　　　It then follows by Prop. (1) that $b' > b_{max}$,
　　　which is impossible by definition.
　Therefore $f_r$ cannot be both bias- and trigger-ambiguous. □

This does not have to mean that ambiguity is completely useless to our purposes however: we may still realize a decrease in recognizability in both the bias and the trigger image by informally relaxing requirements to the level of a *partial ambiguity*, where it suffices that each bias or trigger intensity can lead to more than one of the afterimage intensities.

This idea is implemented in rule set $f_3$ which, while using the same two bias intensities as rule set $f_2$, will target three instead of two afterimage intensities (see Figure 6). We begin its construction by attempting to find a $t_1$ and $t_2$ with $f_a(0, t_1) = f_a(1, t_2)$. Trying to identify such a match can be done using the same procedure as was followed when constructing $f_2$—now, while varying an additional parameter $d$: $t_1 = 0.5 - d$ and $t_2 = 0.5 + d$, with $d$ being raised from zero, thereby moving the trigger intensities away from a middle grey. While doing this, the results produced near $d = 0.13$ appeared to have the greatest stability, yielding $a_2 \approx f_3(0, 0.37)$ and $a_2 \approx f_3(1, 0.63)$.

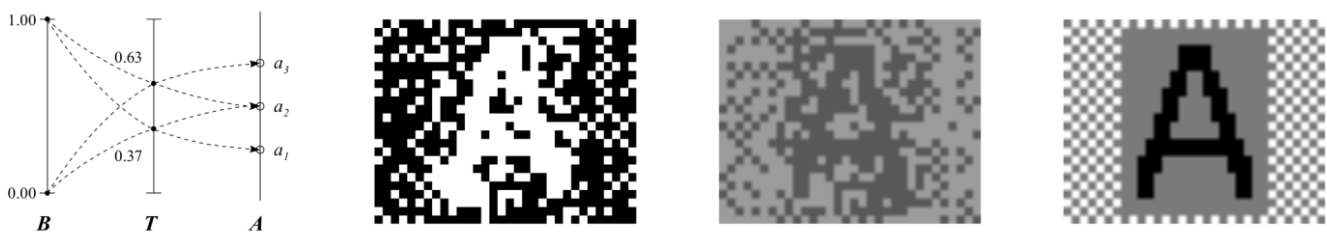

**Figure 6.** Partial ambiguity. Left: diagram of rule set $f_3$, with arrows linking the intensities that are used. Middle, left and right: the randomly distorted bias and trigger images from an image sequence generated by rule set $f_3$. Right: the target afterimage pattern, with $a_1$, $a_2$ and $a_3$ shown as increasingly bright greys.

Although the above has, so far, only provided us with the medium afterimage intensity $a_2$, we can now quickly finalize $f_3$ into a partially ambiguous rule set by having $a_3 = f_3(0, 0.63)$ and $a_1 = f_3(1, 0.37)$.

Now completed, rule set $f_3$ does seem to create a risk regarding recognizability when considering its afterimage intensities $a_1$ and $a_3$: Tracing these back within the diagram of Figure 6, we can see how sections of the afterimage pattern that use these intensities will reoccur, in unambiguously differing greyscale levels, in both the trigger and bias images. However, given that the other areas of these images, leading to $a_2$, will use the same greyscale values randomly, recognition will still be hampered. In Figure 6 and Video Example IV, this is demonstrated using a three-tone variant of the previously used example afterimage pattern.

2.4.2. Scrambling Rule Sets

A basic property of human visual perception is that adjacent areas of a similar shade tend to be grouped together and perceived as a shape. We will now introduce another approach to defining rule sets, using this tendency in a subversive manner. As before, the goal is to have shapes recognized in the afterimage not be recognized by normal viewing of the image sequence producing the afterimage.

First, we will need a tool to look at how rule sets reorder intensities, when comparing the bias images they generate to the target afterimage patterns. Given a rule set $f_r$, we can enumerate the set $B$ of its bias intensities (defined in Section 2.4.1) according to $b_1 < \ldots < b_{|B|}$—just as we have done for the set $A$ of afterimage intensities from the outset. We then define $f_r$'s *mapping scheme* as a tuple of $|B|$ subsets from the set $\{1, \ldots, |A|\}$, where the $i$-th subset consists of all $j$ for which $\exists\, t \in I : f_r(b_i, t) = a_j$.

This means that, reading a mapping scheme from left to right, we find for each bias intensity, from dark to light, the ranks of the afterimage intensities to which it is linked. For example, the mapping scheme for rule set $f_1$ is given by ($\{2\}$, $\{1\}$), meaning that firstly, its darkest bias intensity is mapped to its lightest afterimage intensity; and that secondly, its lightest bias intensity is mapped to its darkest afterimage intensity (see the diagram of Figure 4). As another example, the mapping scheme for rule set $f_2$ is ($\{1, 2\}$, $\{1, 2\}$); just as it will be for any other bias-ambiguous rule set which uses two bias and two afterimage intensities (see the diagram of Figure 5).

Now, suppose we have a target afterimage pattern displaying a shape on a background, with each in one intensity. Both will be repeated in the bias image, in bias intensities according to the rule set used. Arbitrarily choosing one of the bias intensities present for the shape, suppose that it is nearer in brightness to one of the bias intensities present for the background than to all other intensities also present for the shape. Where adjacent, these two bias intensities will tend to visually group together, distorting the perception of the original shape. A rule set $f_r$ will have this property for all possible combinations of uniformly tinted shapes if:

$$
\begin{aligned}
f_r(b, t) = a \Rightarrow \forall a' \in A, a' \neq a : \exists\, b', t' \in I : \\
[\, f_r(b', t') = a' \land \forall b'', t'' \in I, b'' \neq b : \\
f_r(b'', t'') = a \Rightarrow |\,b - b'\,| < |\,b - b''\,|\,]
\end{aligned} \quad (3)
$$

However, given a rule set $f_r$, it may be more intuitive to look at its mapping scheme: if in it, multiple occurrences of the same afterimage intensity are always separated by occurrences of all other afterimage intensities lying strictly inbetween, the above property will hold. Consider for example the mapping schemes ($\{1\}$, $\{2\}$, $\{1\}$) and ($\{1\}$, $\{2\}$, $\{1\}$, $\{2\}$). (For an example illustrating the use of ($\{1\}$, $\{2\}$, $\{1\}$, $\{2\}$), please preview how the rule set shown in the diagram of Figure 8 links its trigger intensities to its afterimage intensities.) Both of the above mapping schemes correspond to rule sets satisfying the property, and both have a scrambling effect on the shapes of the afterimage, as is illustrated in Figure 7. However, the effect of the first mapping scheme is crippled by the fact that it has afterimage intensity $a_2$ preceded by bias intensity $b_2$ only, which greatly aids shape recognition. We still need to make explicit that a rule set $f_r$ should produce each afterimage intensity using multiple bias intensities:

$$
f_r(b, t) = a \Rightarrow \exists\, b', t' \in I, b' \neq b : f_r(b', t') = a \quad (4)
$$

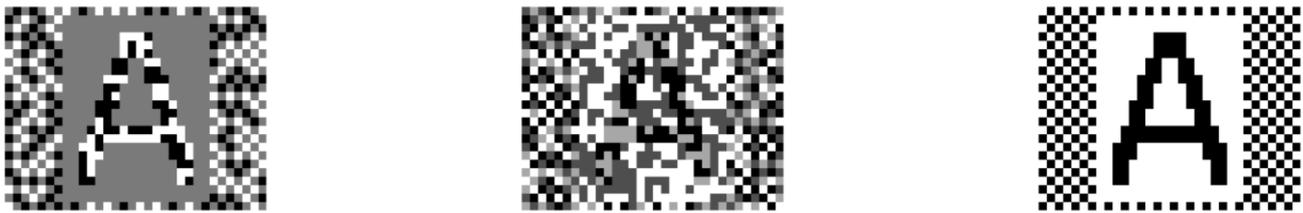

**Figure 7.** Scrambling effects. Right: the target afterimage pattern. Left: a corresponding bias image using the mapping scheme ({1}, {2}, {1}). Middle: a stronger result using mapping scheme ({1}, {2}, {1}, {2}).

Having Props. (3) and (4) now at our disposal, we consider a rule set $f_r$ that satisfies both of them to be *bias-scrambling*. We can then arrive at a very similar definition of *trigger-scrambling* rule sets (and trigger intensity mapping schemes) simply by reiterating the above discussion while replacing bias images with trigger images.

The next example rule set will then satisfy the scrambling requirement for both the bias and trigger images that it generates. These will each be constructed in four different intensities, starting with black and white bias intensities that are combined in such a way with two trigger intensities near middle grey as to produce two afterimage intensities that are ordered the other way around: $a_1 = f_4 (1, 0.52)$ and $a_2 = f_4 (0, 0.48)$, and see Figure 8. To this we then add two inner bias intensities and two outer trigger intensities, by searching for two pairs of identical intensities that will also yield $a_1$ and $a_2$ in the afterimage. Here, iteratively testing with all-$a_1$ and all-$a_2$ patterns (like before in Section 2.4.1) seemed to give the most stable results when choosing $a_1 \approx f_4 (0.39, 0.39)$ and $a_2 \approx f_4 (0.62, 0.62)$.

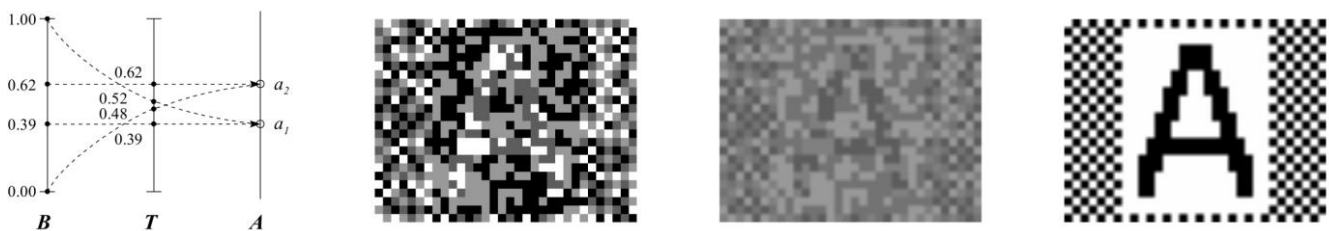

**Figure 8.** Bias-scrambling, trigger-scrambling. Left: diagram of rule set $f_4$, with arrows linking the intensities that are used. Middle, left and right: the scrambled bias and trigger images from an image sequence generated by rule set $f_4$. Right: the target afterimage pattern, with $a_1$ shown as black and $a_2$ shown as white.

This means that both 0.39 and 0.62 will play double roles in $f_4$: on the one hand, as the inner bias intensities, resulting in the bias intensity mapping scheme ({2}, {1}, {2}, {1}); and on the other hand, as the outer trigger intensities, resulting in the trigger intensity mapping scheme ({1}, {2}, {1}, {2}).

Visual outcomes of this are demonstrated by the image sequence of Video Example V, also presented in Figure 8. When trying to assess how well the afterimage pattern is being scrambled, we can see that this is impaired in the trigger image by something documented in the rule set diagram: Each of the outer trigger intensities will visually group together with both of the inner trigger intensities almost equally well, due to their relative closeness. Placing the inner trigger intensities further apart involves a trade-off, however, because this in turn will place the afterimage intensities closer together, and thereby reduce afterimage contrast.

2.4.3. Hybrid Rule Sets

Reflecting on ambiguity versus scrambling, it is now clear that both have inherent strengths and weaknesses: Ambiguity is better for really guaranteeing that the afterimage pattern will not be recognizable—but it restricts this guarantee to either the bias or the trigger image. Scrambling, as we have seen, does allow simultanous use in the bias and trigger images—but it cannot provide a hard guarantee of unrecognizability. Given these characteristics, it becomes worthwhile to explore whether the two approaches could be combined within a single rule set.

The construction of our first example of this can start from the middle two trigger intensities of rule set $f_4$ (see Figure 8), but with each intensity placed somewhat further away from medium grey. This for a rule set $f_5$ initially defined by $a_1 = f_5 (1, 0.57)$ and $a_2 = f_5 (0, 0.43)$. Searching for pairs of same-value bias and trigger intensities that would also produce these afterimage intensities yielded $a_1 \approx f_5 (0.43, 0.43)$ and $a_2 \approx f_5 (0.57, 0.57)$. This then completes a rule set which, like rule set $f_4$, on the one hand is bias-scrambling, using four bias intensities with a bias intensity mapping scheme of ({2}, {1}, {2}, {1}); but which, on the other hand, also is trigger-ambiguous for its two trigger intensities. This is illustrated in Figure 9, along with an image sequence generated using $f_5$ that is demonstrated in Video Example VI.

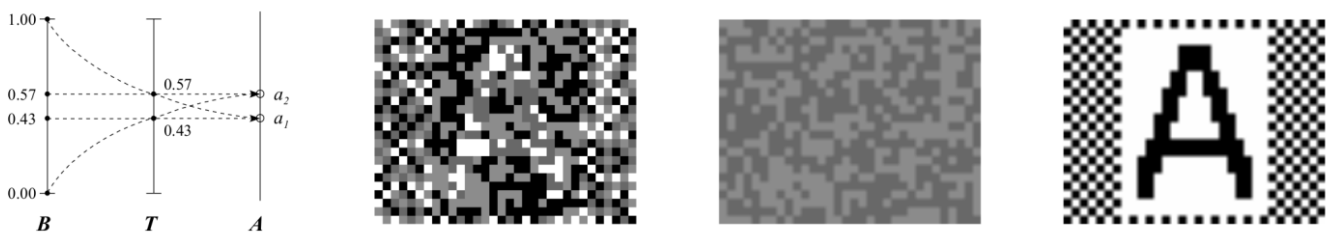

**Figure 9.** Bias-scrambling, trigger-ambiguous. Left: diagram of rule set $f_5$, with arrows linking the intensities that are used. Middle, left: the scrambled bias image from an image sequence generated by rule set $f_5$. Middle, right: the random trigger image. Right: the target afterimage pattern, with $a_1$, $a_2$ shown as before. Please note that the actually experienced afterimage may be quite different: see the text.

It appears, however, that in rule set $f_5$, $a_1$ and $a_2$ are too near eachother to obtain a good contrast in the afterimage: With the resulting greys seeming quite similar, shape is hard to separate from background. To improve on this, construction of our next rule set will ensure from the start that an afterimage contrast comparable to that of $f_4$ will again be obtained—while switching the scrambling and ambiguity requirements.

Constructing rule set $f_6$ starts by copying from $f_4$, and initially defining its afterimage intensities as $a_1 = f_6 (1, 0.52)$ and $a_2 = f_6 (0, 0.48)$. Then, while testing iteratively using all-$a_1$ and all-$a_2$ afterimage patterns, we look for a $t_1$ that yields $a_1 \approx f_6 (0, t_1)$ and a $t_4$ that yields $a_2 \approx f_6 (1, t_4)$. The most stable impression here seemed to result when choosing $a_1 \approx f_6 (0, 0.25)$ and $a_2 \approx f_6 (1, 0.74)$. This then completes a rule set that is trigger-scrambling, based on four trigger intensities with a corresponding mapping scheme of ({1}, {2}, {1}, {2}). At the same time, the two bias intensities satisfy bias ambiguity. This is illustrated in Figure 10, along with an image sequence generated using $f_6$ that is demonstrated in Video Example VII.

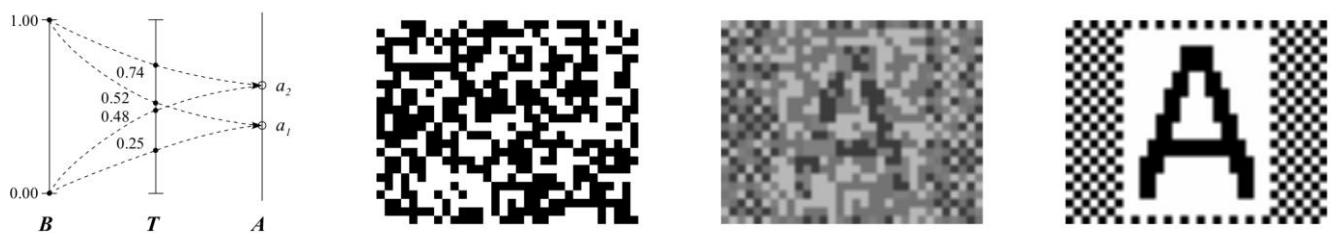

**Figure 10.** Bias-ambiguous, trigger-scrambling. Left: diagram of rule set $f_6$, with arrows linking the intensities that are used. Middle, left: the random bias image from an image sequence generated by rule set $f_6$. Middle, right: the scrambled trigger image. Right: the target afterimage pattern, with $a_1$, $a_2$ shown as before.

Comparing the current rule set to its predecessor, even though $f_6$ does improve on the contrast between afterimage intensities, $f_5$ will probably produce the more effective scrambling: Its middle bias intensities are relatively wider apart than the middle trigger intensities of $f_6$, which suffers even more from the same weakening effect on scrambling that has already been discussed for rule set $f_4$.

2.4.4. Multi-Trigger Sequences

By their definition, rule sets with bias ambiguity enable a random, but already fixed bias image to still target any possible pattern in the afterimage. This can now be tested using the bias-ambiguous, trigger-scrambling rule set $f_6$: As is not possible when mirroring the same idea using trigger ambiguity, in a single run, one pronounced bias impression generated by $f_6$ might be combined with more than one trigger image to yield the perception of multiple afterimage patterns.

One way of implementing this is to simply show two successive trigger images instead of one. This has been done in Video Example VIII, illustrated in Figure 11, successively targeting the words "hello" and "world" in the afterimage. Here, the initial trigger image is shown for 1.5 s, in a tentative compromise: Longer durations would have given the viewer more time to recognize the initial afterimage shape, while however also diminishing the quality of the second afterimage; and shorter durations would have had the opposite effect. In another trade-off, showing the bias image for progressively longer time periods would on the one hand help here—by yielding a better contrast in especially the second afterimage—but it would also decrease viewing comfort and afterimage stability. The tentative compromise for bias viewing duration was then to increase it by 10 s, to 30 s. The above parameters, like the others before them, can be freely experimented with using the software in the Electronic Appendix.

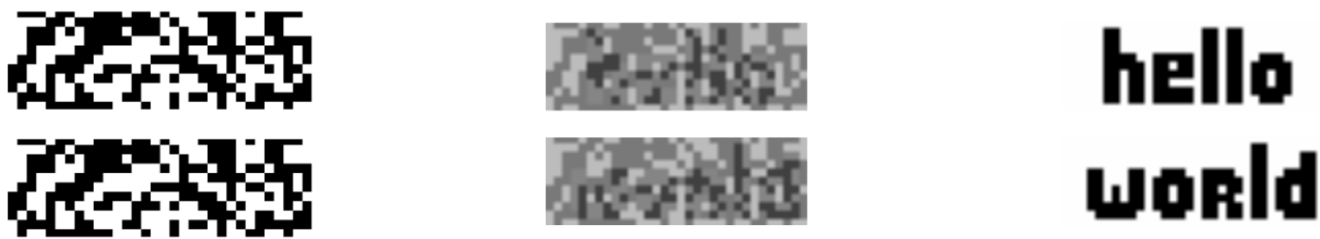

**Figure 11.** Multiple trigger images for a "hello world" example. A sequence generated by rule set $f_6$. Left: the middle section of the bias image (shown twice). Middle: the corresponding sections of the first and second trigger images. Right: the target afterimage patterns ($a_1$, $a_2$ as before). This sequence also demonstrates the effects of scrambling when using finer target shapes.

## 3. Results of Experimental Evaluation

Before the main experiment that will be discussed below, the approach was tested in a pilot experiment. Here, two image sequences were used that were each generated from, as their afterimage targets, word shapes consisting of 5 letters. This was done using rule sets $f_3$ and $f_6$, respectively. Showing these image sequences to 5 test subjects resulted in none of them recognizing any word shapes in the separate bias and trigger images, and all of them reporting the target 5-letter word shapes in the afterimage effect.

*3.1. General Overview and Construction*

Like the pilot, the follow-up main experiment was a controlled experiment, similarly comparing afterimage viewing to normal viewing. Some key differences, however, were a greater number of test subjects ($n$ = 14), a balanced two-samples rather than a paired design, and a greater emphasis on randomization.

The aim of the experiment was to assess whether the automated approach presented in Section 2 is indeed capable of displaying, via the retinal afterimage, predefined 2D shapes which are verified to not be recognized by test subjects due to normal viewing. The test subjects who took part in the experiment were of course naive to the target patterns, as well as to the visualization method itself.

Regarding visual fixation and rasterization, the visualization method was tentatively parametrized according to the outcomes discussed in Sections 2.1 and 2.2. The next question was then to choose such rule sets from Section 2.4 as would be representative for the approach as a whole. The first two rule sets $f_1$ and $f_2$, being trigger- and bias-ambiguous, respectively, clearly allow for easy visual recognition of the target shape, in the images where ambiguity does not apply. Therefore, these rule sets would not be used. Then comparing bias- and trigger-scrambling rule set $f_4$ to rule set $f_6$, the latter, while also being trigger-scrambling, is bias-ambiguous, thereby guaranteeing unrecognizability where $f_4$ cannot do so. Rule set $f_6$ was therefore preferred over $f_4$, and subsequently, also over $f_5$, the bias-scrambling, trigger-ambiguous rule set, because $f_6$ was designed to induce a stronger contrast in the retinal afterimage than that of $f_5$. Finally, since the remaining rule set, partially ambiguous $f_3$, did not seem necessarily worse than $f_6$, both were selected for use in the experiment.

The next decisions were then about what kind of target pattern to try to visualize via the retinal afterimage. In any case, to avoid distraction, backgrounds of a single uniform target level would be used. The foreground for each afterimage would then consist of a single word, since words are a familiar kind of 2D shape that can be rapidly recognized. Informed by a result that during regular viewing, strings of letters are more rapidly recognized as words when placed in the center of visual fixation rather than in its periphery [27], these word shapes would then be placed around the fixation crosshair.

Within the shapes, the lines used were mostly two square units thick. This is because it became apparent, when using lines one unit thick, that the distortion of individual letters would sometimes make them unreadable in the afterimage (see also Section 2.2).

Using this type of stimuli, the general idea was then to create two experimental conditions, differing only on very specific points. In both conditions, test subjects would view the exact same visual stimuli, i.e., pairs of bias and trigger images. However, under the control condition, these image pairs would be viewed in a normal, everyday fashion; while under the alternative condition, they would be viewed in such a way as to induce an afterimage effect. From here on, these two conditions will be referred to as *normal viewing* and *afterimage viewing*. Under afterimage viewing, each bias image naturally was followed by its trigger image; while under normal viewing, the trigger image was shown first, then the corresponding bias image. The order was reversed here to avoid the possibility of accidental visual fixation actually effectively yielding the other experimental condition.

Another important design choice was that test subjects would be split into two equally sized subgroups, with each subgroup being exposed to only one of the experimental conditions. This in order to help secure the independence of measurements, by eliminating any possibility that visual recognition by a subject, on exposure to a specific image pair, might somehow be influenced by having been exposed to the same image pair before, under a different experimental condition.

Given that test subjects would be completely new to the afterimage viewing method, learning effects were to be expected for it, and therefore, initial viewing dry runs were included in the experiment. Similarly but separately, subjects would also be new to attempting visual recognition during the experimental afterimage effect itself, and therefore, additional recognition dry runs were included as well. Given that preliminary experimentation had indicated that 3-letter word shapes were harder to recognize in the afterimage than 5-letter word shapes, the former were used in recognition dry runs, and the latter in actual experimental runs. Both in recognition dry runs and experimental runs, the order in which different image pairs were shown was randomized for each individual test subject.

*3.2. Procedure*

The 14 test subjects for the experiment were 16 to 29 years of age, 3 male, 11 female. All were new to the visual stimuli and to the afterimage viewing method in general. Due to near-sightedness, far-sightedness and astigmatism, 9 test subjects needed correction by glasses or contact lenses. For the remaining 5, this was not the case. One subject reported having dyslexia. All reported not being color blind; not having other visual recognition issues; and having been comfortable during the experiment. All test subjects except one reported being able to concentrate well during the experiment, while two reported being not well-rested.

The Electronic Appendix, which is viewable using any standard browsing software, offers, among other things, playback of the images and image sequences used in the experiment. It was held indoors during daytime, with no artificial light, and indirect daylight. In general, the experimental circumstances were kept similar as in Section 2, although now, a MacBook Pro laptop (Retina, 13-inch, early 2015) running macOS 12.2.1 was used to show the image sequences. Its display was set to maximum brightness and yielded an ~ 18 by 13 cm display area for the experimental images, which were viewed from a distance of ~ 50 cm. Separate images were each shown for 20 s; and during the sequences, a bias image was always shown for 20 s and then followed immediately by the trigger image, shown for 5 s.

For the test subjects assigned to afterimage viewing, during the initial viewing dry runs, the third, chequered sequence shown in Video Example I was used to familiarize them with the visualization method (see also Figure 3). Once seated in front of the display, the subjects were informed that they would be shown a series of two images, with the first changing into the second. They were asked to concentrate on the instant when this change would occur, and instructed beforehand to place their elbows on a stable surface; to rest their head on both hands; and then, throughout the sequence, to focus visually on the center of the crosshair. When the sequence had finished, they were asked a question which would return again and again during the rest of the session: Had they recognized a pattern, and if so, what was it? Also, the test subjects were alerted to how the light grey fuzzy squares preceded by black ones seemed to light up during the sequence, which was then repeated one or more times until the test subjects signalled that they had gotten used to the effect.

For the test subjects assigned to normal viewing, during the initial viewing dry run, the separate trigger image of the same chequered sequence was shown, and then its bias image. The test subjects were asked to look at the stimuli as if at ordinary photographs or illustrations. After this, the question on recognition was asked.

Beyond this point in the experiment, image pairs were presented using the same procedures, only now the subjects wrote down their answers in a form, and did not receive feedback anymore. They could however indicate uncertainty about their response, in which case an image pair was repeated until a final answer was settled upon.

Both under normal viewing and afterimage viewing, test subjects then next performed the recognition dry runs. These, in an individually randomized order, presented two image pairs: One generated by rule set $f_3$, based on a 3-letter pattern spelling "red"; and another generated by rule set $f_6$, based on a 3-letter pattern spelling "low". After these were completed, the test subjects had the opportunity to write down any general remarks, and there was a short break. Then, finally, the experimental runs also presented two image pairs in a randomized order: One generated by rule set $f_3$, based on a 5-letter pattern spelling "light"; and another generated by rule set $f_6$, based on a 5-letter pattern spelling "hello". After this, the answers to some additional questions could be noted down in the form, and the experimental session was concluded.

*3.3. Results*

As has been described above, during the experiment, both the test subjects subject to normal viewing and those subject to afterimage viewing were asked qualitative questions about each of the stimuli just after their presentation.

Under *normal viewing*, this resulted in none of the test subjects ever reporting the recognition of a word pattern, let alone reporting recognition of the target afterimage pattern. Subjects would report seeing darker and lighter areas, smudges, protrusions, or shapes like a "QR code". Perhaps most imaginatively, one subject reported seeing "trails of viscous cheese sauce". In any case, the observation coming closest to word pattern recognition was that a single subject reported seeing an area somehow seeming "textual" in the $f_3$-generated image pair. Notably, this was also the only subject who reported having dyslexia.

Under *afterimage viewing*, on the other hand, all test subjects always did report seeing word shapes—except for one outlier, who did not report word shapes for any of the image pairs. Reports by this subject of "dark grey roundish shapes" and "waving lines from lower left to upper right" seem to indicate that something may have gone wrong with maintaining visual fixation, after most letters had been recognized correctly during the earlier recognition dry runs. Overall, except for the outlier, all test subjects correctly recognized the target afterimage pattern for the $f_3$-generated image pair: Here, a word shape spelling "light" was reported 6 out of 7 times. For the $f_6$-generated image pair, a word shape spelling "hello" was correctly reported 4 out of 7 times.

As was already expected from the results of the earlier pilot experiment, the clear contrast present in qualitative reporting described above, between no recognition versus often correct recognition, provided a good basis for a further quantitative comparison between the two experimental conditions of *normal viewing* versus *afterimage viewing*.

For this comparison, first, the data from the filled-in questionnaires was converted from qualitative to quantitative for each test subject, using the single measure of "*% of times the afterimage pattern was correctly recognized in the image pair*". It should be appreciated that this is a relatively strict measure, given that we do not count correctly recognized individual letters or word segments: Anything less than correct recognition of the complete word counts as zero. This is further strengthened by the fact that under normal viewing, a counted instance of "the afterimage pattern was *not* correctly recognized" actually means "*no word pattern* was reported *whatsoever*". A final factor further increasing the strictness of this measure is the fact that the rule sets were tested on a different display than that for which they had been fine-tuned, leading to suboptimal afterimage visualization.

Overall—i.e., for both of the tested rule sets taken together, and even when also including the data from the outlier test subject—this then yielded an observed difference in group means of 71% under afterimage viewing relative to 0% under normal viewing. To assess the reliability of this large observed difference, a *t*-test was used. Performing a two-samples, heteroskedastic *t*-test yielded *p* = 0.0030. Figure 12 gives an overview of the results of the experiment.

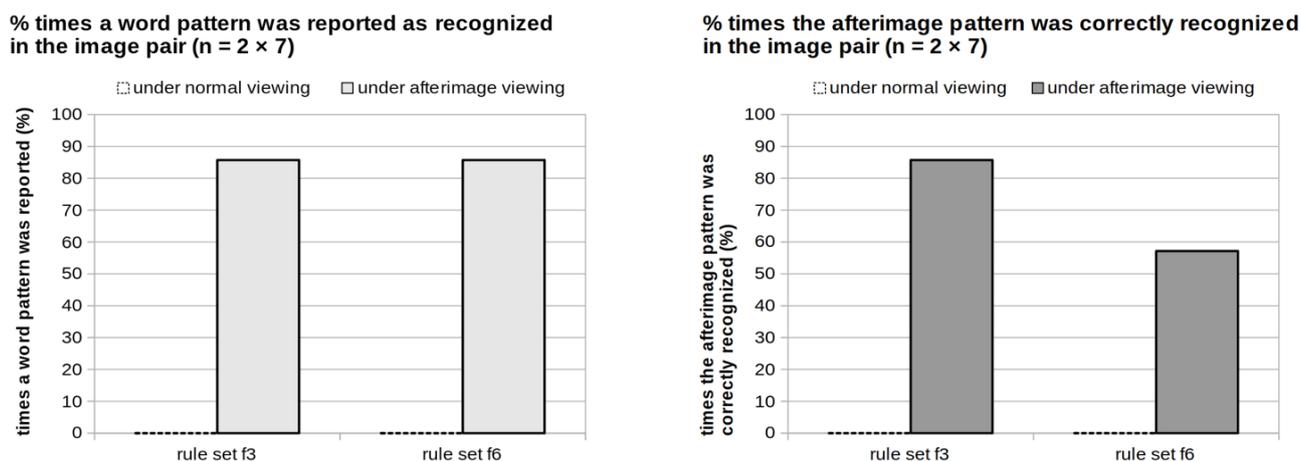

**Figure 12.** Results of the experiment: word shape recognition under normal viewing versus under afterimage viewing. *Left and right*: under normal viewing, none of the test subjects ever reported recognizing any word pattern, let alone the correct target afterimage patterns. *Left*: under afterimage viewing, only a single test subject did not report recognizing word patterns. This was the same test subject for both rule sets $f_3$ and $f_6$. *Right*: all other test subjects correctly recognized the target afterimage pattern when using rule set $f_3$.

## 4. Discussion

As was mentioned before, preliminary testing before the experiment indicated that 3-letter word shapes were harder to recognize in the induced afterimage than 5-letter word shapes. This might be explained by word shapes of 3 letters being less robust against irregularities during afterimage display (see Section 2) than shapes consisting of 5 letters.

If so, this could motivate the development of a more sophisticated working model of how perceived afterimage intensity is induced, and of an improved rasterization method based on the model. Such development could be based on results from the literature, including [23] which, for structured afterimages that completely or partially disappear and reappear, studies what factors may influence this; as well as [24], which studies how selective attention influences this process, including how enclosed regions become filled-in; and also [25], which studies how the geometrical contours of an afterimage (induced beforehand) aligning retinally with the contours in an image being subsequently viewed may lead to an afterimage that appears more stable and more intense; and also [28], which studies the influence of post-adaptation contours on effects of color spreading and filling-in.

During adaptation as well as after, the phenomenon of induced simultaneous contrast influences how colors and greys appear to the viewer [26]. This, too, should be considered explicitly when further refining how to computationally induce the retinal afterimage. Adaptation itself could be made more rapid by using surroundings that are darker and/or stimuli that are brighter than those used here. To allow for eye movements that are more natural during phases of adaptation, the output via visual display technologies could become closely based also on real-time input from eye tracking technologies.

## 5. Future Work

An important point to make, however, is that it cannot be within the scope of this paper to determine which possible steps for further development of the approach should be taken first, and in what order. This will depend on how visual artists will choose to view the retinal afterimage as a *medium*: e.g., primarily as a means for *illusory display*; as an approach yielding *idiosyncratic visual experiences*; or as an *addition to existing forms of computer graphics*.

For example, the approach presented in Section 2 can be readily expanded in order to also induce colored afterimages. Here, if the artist in question is interested in conveying information by purely illusory means, it will be important to start from and expand on the formally defined and empirically tested constraints that enabled display exclusive to the retinal afterimage. However, if the same artist primarily is interested in working with the *visual experience* of perceiving shapes and patterns via the retinal afterimage, it may be smarter to simplify any immediate further development of the approach by first relaxing the formal constraints for exclusive display, e.g., beyond what was done at the end of Section 2.4.1 and in Sections 2.4.2 and 2.4.3.

Another good example here is the choice whether or not to immediately add visual fixation as an *input* when further developing the approach. If a visual artist intends to work using fixed pairs of bias and trigger images, it may suffice to replace the fixation crosshair by a visual element that triggers fixation in the context of the art work [29]. However, if an artist intends to work with multiple triggers that each may support visual fixation, monitoring the series of actually occurring fixations may become desirable, and this implemented within the work itself. As has been demonstrated in Section 2.4.4, a visual artist may apply bias-ambiguous rule sets to enable the perception of different possible afterimage shapes using a single, random initial bias image. The choice here may be left to the viewer, who by eye movement and subsequent fixation alone might select the actual trigger image from a number of simultaneously offered alternatives. Here, if display technology with integrated eye tracking would be used, the visualization software could keep track of these choices, as they are being made, thereby enabling the visual artist to prepare an overall afterimage experience for the viewer that is interactive.

Also, regardless of whether the visual artist will intend to use the retinal afterimage itself verbally or non-verbally, figuratively or abstractly—it may well be *in combination with existing forms of computer graphics*. If this is the case, it may be an effective choice to continue development of the approach using the display technologies of Virtual Reality (VR) and Augmented Reality (AR) devices, as these will allow exploring addition of the retinal afterimage to a wide range of different forms of interactive 3D computer graphics. A common starting point here may often be the careful texture mapping of bias and trigger images into 3D environments.

In light of the earlier points made above, it is then especially relevant to also note that high-end AR/VR headsets with integrated eye tracking have recently become commercially available [30].

## 6. Conclusion

Finally, more immediately reflecting on the outcome of the experiment described in Section 3, its results indicate that the computational method we have developed can be used to successfully induce a subset of all specifiable, recognizable 2D shapes *specifically* via the retinal afterimage. Therefore, and in this sense, the method seems to constitute a positive answer to the research question: Techniques for producing the retinal afterimage can become partly automated. For visual artists interested in the effect this yields the benefit that controlled visual complexity using the retinal afterimage can be created in less time.

The supplementary Electronic Appendix to this paper also contains a publicly licensed software toolkit, provided in source format and running on the Max programming environment in widespread use among contemporary artists [31]. The toolkit can be used to completely reproduce the approach, and to freely experiment with it.

**Supplementary Materials:** The Electronic Appendix is available online at https://www.mdpi.com/article/10.3390/bdcc6030097/s1.

**Author Contributions:** Conceptualization, S.d.J.; methodology, S.d.J.; software, S.d.J.; validation, S.d.J. and G.v.d.V.; formal analysis, S.d.J.; investigation, S.d.J.; resources, S.d.J.; data curation, S.d.J.; writing—original draft preparation, S.d.J.; writing—review and editing, S.d.J. and G.v.d.V.; visualization, S.d.J. All authors have read and agreed to the published version of the manuscript.

**Funding:** This research received no external funding.

**Institutional Review Board Statement:** All subjects gave their informed consent for inclusion before they participated in the study. The study was conducted in accordance with the Declaration of Helsinki of 1975, revised in 2013, and the protocol was approved by the Ethics Review Committee of the Faculty of Science (FWN) of Universiteit Leiden (project identification code: 2022-011).

**Informed Consent Statement:** Informed consent was obtained from all subjects involved in the study.

**Data Availability Statement:** The data presented in this study are included in their raw form in the Electronic Appendix (see Supplementary Materials, above).

**Acknowledgments:** The first author was originally inspired to begin this research by seeing Jeremy Hinton's *Lilac Chaser*, and then reading Michael Bach's online explanation of it. His distinct thanks go also to dr. Walter Kosters, for kindly providing feedback on the formal notation. Many thanks as well to the test subjects, who kindly volunteered without receiving a financial reward. Figure 1b is based on a detail from a wedding photograph taken in the early 1940s, of Roos van Engelen, my dear grandmother, who first tried to show me the retinal afterimage.

**Conflicts of Interest:** The authors declare no conflict of interest.

## References

1. Hering, E. *Outlines of a Theory of the Light Sense*; Harvard University Press: Cambridge, MA, USA, 1964.
2. Hurvich, L.M.; Jameson, D. An opponent-process theory of color vision. *Psychol. Rev.* **1957**, *64*, 384–404. [CrossRef] [PubMed]
3. Rathus, S.A. *Psychology: Concepts and Connections*, 10th ed.; Cengage Learning: Stamford, CT, USA, 2012.
4. Poplawski, P. *Encyclopedia of Literary Modernism*; Greenwood Press: Westport, CT, USA, 2003.
5. Gardner, H.; Kleiner, F.S. *Gardner's Art through the Ages: The Western Perspective, Volume 2*; Cengage Learning: Stamford, CT, USA, 2010.
6. Chevreul, M.E. *De la Loi du Contraste Simultané des Couleurs, et de l'Assortiment des Objets Colorés, Considéré d'Après Cette Loi*; Pitois-Levrault: Paris, France, 1839.
7. Chevreul, M.E. *The Principles of Harmony and Contrast of Colours, and Their Applications to the Arts*; Longman, Brown, Green, and Longmans: London, UK, 1855.
8. Signac, P. *D'Eugène Delacroix au Néo-Impressionnisme*; Éditions de la Revue Blanche: Paris, France, 1899.
9. Scholastic Inc. Georges Seurat—Working with Color. In *Scholastic Art*; Scholastic Inc: New York, NY, USA, 2008; Volume 39, pp. 1–4, Teacher's Edition.
10. Houston, J.; Hickey, D. *Optic Nerve: Perceptual Art of the 1960s*; Merrell Publishers: London, UK, 2007.
11. Seitz, W.C.; Museum of Modern Art. *The Responsive Eye*; Museum of Modern Art: New York, NY, USA, 1965.
12. Sylvester, D.; De Sausmarez, M. *Bridget Riley: Works 1960–1966*; Ridinghouse: London, UK, 2012.
13. Madden, D.; Spike, N.; Spike, J.T. *Anuszkiewicz: Paintings & Sculptures, 1945–2001: Catalogue Raisonné*; Edizioni Centro Di: Firenze, Italy, 2010.
14. Morgan, A.L. *The Oxford Dictionary of American Art and Artists*; Oxford University Press: Oxford, UK, 2007.
15. Parola, R. *Optical Art: Theory and Practice*; Reinhold Book Corporation: New York, NY, USA, 1969.
16. Goethe, J.W. *Bild eines Mädchens in umgekehrten Farben, gezeichnetes Nachbild in Komplementärfarben*; watercolor with Pencil Outlines on Paper; Klassik Stiftung Weimar: Weimar, Germany, 1795/1805.
17. Jenkins, R.; Wiseman, R. Darwin illlusion: Evolution in a blink of the eye. *Perception* **2009**, *38*, 1413–1415. [CrossRef] [PubMed]
18. Chilvers, E.I.; Albers, Josef. *The Oxford Dictionary of Art and Artists (Oxford Reference Online)*; Oxford University Press: Oxford, UK, 2009.
19. Albers, J. *Interaction of Color–Revised and Expanded Edition*; Yale University Press: New Haven, CT, USA, 2006.
20. Brakhage, S. The stars are beautiful. In *Essential Brakhage: Selected Writings on Filmmaking*; McPherson & Company: Kingston, NY, USA, 1967; pp. 134–137.
21. Verheijen, F.J. A simple after image method demonstrating the involuntary multidirectional eye movements during fixation. *Opt. Acta* **1961**, *8*, 309–312. [CrossRef] [PubMed]
22. Martinez-Conde, S.; Macknik, S.L.; Hubel, D.H. The role of fixational eye movements in visual perception. *Nat. Rev. Neurosci.* **2004**, *5*, 229–240. [CrossRef] [PubMed]
23. Wade, N.J. Why do patterned afterimages fluctuate in visibility? *Psychol. Bull.* **1978**, *85*, 338–352. [CrossRef]
24. Lou, L. Effects of voluntary attention on structured afterimages. *Perception* **2001**, *30*, 1439–1448. [CrossRef] [PubMed]
25. Daw, N.W. Why after-images are not seen in normal circumstances. *Nature* **1962**, *196*, 1143–1145. [CrossRef] [PubMed]
26. Anstis, S.; Rogers, B.; Henry, J. Interactions between simultaneous contrast and coloured afterimages. *Vis. Res.* **1978**, *18*, 899–911. [CrossRef]
27. Lee, H.-W.; Legge, G.E.; Ortiz, A. Is word recognition different in central and peripheral vision? *Vis. Res.* **2003**, *43*, 2837–2846. [CrossRef]
28. Van Lier, R.; Vergeer, M.; Anstis, S. Filling-in afterimage colors between the lines. *Curr. Biol.* **2009**, *19*, R323–R324. [CrossRef] [PubMed]
29. Kirtley, C. How images draw the eye: An eye-tracking study of composition. *Empir. Stud. Arts* **2018**, *36*, 41–70. [CrossRef]

30. Microsoft Corporation. Eye Tracking on HoloLens 2. Available online: https://docs.microsoft.com/en-us/windows/mixed-reality/design/eye-tracking (accessed on 22 August 2022).
31. Puckette, M. Max at seventeen. *Comput. Music J.* **2002**, *26*, 31–43. [CrossRef]